\documentclass[preprint,showpacs,preprintnumbers,amsmath,amssymb,pra,aps]{revtex4-1}
\usepackage{graphicx,textcomp}
\usepackage{dcolumn}
\usepackage{bm}
\usepackage{amsmath}
 \usepackage{color}

\begin{document}

\title{Photoassociation of a cold atom-molecule pair: long-range quadrupole-quadrupole interactions.}
\author{M. Lepers$^{1}$, O. Dulieu$^{1}$, V. Kokoouline$^{1,2}$}
\affiliation{$^{1}$Laboratoire Aim\'e Cotton, CNRS, UPR3321, B\^at. 505, Univ Paris-Sud, 91405 Orsay Cedex, France\\
$^{2}$Department of Physics, University of Central Florida, Orlando, Florida 32816, USA }
\email[M. Lepers: ]{maxence.lepers@u-psud.fr}

\begin{abstract}
The general formalism of the multipolar expansion of electrostatic interactions is applied to the calculation of the potential energy between a excited atom (without fine structure) and a ground state diatomic molecule at large mutual separations. Both partners exhibit a permanent quadrupole moment, so that their mutual long-range interaction is dominated by a quadrupole-quadrupole term, which is attractive enough to bind trimers. Numerical results are given for an excited Cs($6P$) atom and a ground state Cs$_2$ molecule. The prospects for achieving photoassociation of a cold atom/dimer pair are thus discussed and found promising. The formalism can be generalized to the long-range interaction between molecules to investigate the formation of cold tetramers.
\end{abstract}

\pacs{31.30.jh, 67.85.-d}

\maketitle

\section{Introduction}

Since it was proposed by Thorsheim \textit{et al.} \cite{thorsheim87} in 1987, and first observed for Sodium \cite{lett93} and for Rubidium \cite{miller93} atoms in 1993, the photoassociation (PA) of pairs of ultracold atoms has had a tremendous impact on research in atomic, molecular, and optical physics at low temperatures. There are several recent review articles devoted to the various aspects of PA \cite{lett95,stwalley99,jones06,dulieu09}, therefore we briefly recall below some of the main features of the PA process, which gave rise to a new high-resolution spectroscopic technique, \textit{i.e.} the PA spectroscopy.
Due to their extremely low relative kinetic energy, atoms from an ultracold gas can be associated via a quasi-resonant free-bound dipolar transition to form an electronically excited molecule, which is often created in a highly excited rovibrational level. As the PA process is mainly controlled by the long-range electrostatic interactions between cold atoms, it has been used as a high-resolution spectroscopy technique for highly rovibrational levels. The highly excited rovibrational levels observed using PA correspond to vibrational motion of a molecule with much larger extension than the usual chemical bond \cite{jones96,wang97,fioretti99,comparat00,fioretti01}. Such molecules with a very large amplitude of vibration had been  predicted fifteen years before the mentioned experiments \cite{movre77,stwalley78}. The spectroscopy of the highly excited rovibrational levels of photoassociated dimers permitted, in particular,  to determine the most accurate values of the radiative lifetime of the first excited state of alkali-metal atoms (see for instance \cite{bouloufa09}). Another example of PA application is the formation of stable ultracold molecules, reported initially for Cs$_2$ \cite{fioretti98}, and later for many other homonuclear and heteronuclear alkali-metal diatomic molecules \cite{nikolov99,gabbanini00,fatemi02,wang04,haimberger04,kerman04,mancini04,kraft06}.

With the improvement of the experimental techniques at ultra-cold temperatures, the study of the quantum dynamics of few-body systems in the ultracold regime has become possible, as illustrated by the recent observations of cold collisions between atoms and molecules \cite{cvitas05,zahzam06,staanum06,hudson08,knoop10}. Such phenomena attract at present a lot of interest as they represent the first manifestation of a novel ultracold chemistry, which is controlled by the quantum nature of the colliding partners \cite{ospelkaus10,ni10}. In particular, at certain conditions, the ultracold few-body dynamics exhibits universal (i.e. species-independent) properties for long-range bound states and resonances (see, for example, Ref. \cite{mehta09,levinsen09} and references therein), nowadays referred to as the Efimov physics \cite{efimov70,efimov10}. The Efimov states have recently been observed experimentally \cite{kraemer06,pollack09,zaccanti09}.

All these developments concern atoms and molecules in their electronic ground state. The purpose of the present study, as the first of a series of papers, is to investigate the next step towards ultracold chemistry: the association of ultracold atoms and molecules with a laser field to create weakly bound trimers or tetramers in an excited electronic state, which has not been previously discussed in the literature to our knowledge. Just like for  pairs of atoms, the PA probability is determined by the long-range interactions between the colliding partners. Here, we consider the long-range interaction between a $^1\Sigma_g^+$ molecule in a given rovibrational level $(v_d, j)$ with an atom in a $P$ electronic level without fine structure. This situation will be illustrated with the interaction between a ground state Cs$_2$ molecule, and an excited Cs($6P$) atom. The leading term of this interaction at large interparticle distances $R$ is a quadrupole-quadrupole term varying as $R^{-5}$, but can be easily generalized to other species. The present work can also be viewed as a step beyond several related studies. The quadrupole-quadrupole interaction between two exited $^2P$ atoms has been calculated for alkali-metal atom pairs \cite{marinescu97} and for the LiB molecule \cite{pouchan96}. In Refs. \cite{rerat03,merawa03}, the van der Waals interaction (varying as $R^{-6}$) between alkali-metal dimers in the $(v_d=0, j=0)$ level their lowest triplet state and a ground state alkali-metal atom has been determined, while in Refs. \cite{bussery08,bussery09} the interaction between a $^2\Pi$ molecule and a $^3P$ atom \textit{at fixed geometries} is obtained as a sum of a dipole-quadrupole term (in $R^{-4}$) and a quadrupole-quadrupole term (in $R^{-5}$).

In Section \ref{sec:multipole}, we briefly review the main ingredients of the perturbative approach based on the multipolar expansion of the long-range interaction between the two fragments. Section \ref{sec:c5} is devoted to the calculation of $C_5$ coefficients of the long-range behavior of molecular potentials. We consider the general case of an arbitrary rotational state $j$ of the dimer as well as we give an analytical solution for the particular case of $j=1$.  Atomic units (a.u.) for distances (1 a.u. = 0.0529177~nm) and for energies (1~a.u. = 219474.63137~cm$^{-1}$) will be used throughout the paper, except otherwise stated.

%The initial state could be either three unbound atoms, an atom-molecule pair, or a pair of molecules, and the PA probability will depend on their quantum properties, \textit{i.e.} the presence of a Feshbach resonance or of an Efimov resonance. The enhancement of PA probability for atom pairs due to a Feshbach resonance has been recently predicted and labeled "Feshbach -Optimized Photoassociation" \cite{pellegrini08}.

\section{Interaction potential and perturbation theory}
\label{sec:multipole}

We start the description of the present theory from the general case, as for instance in Refs. \cite{flannery05,bussery08}. We consider two charge distributions, $A$ and $B$, far from each other such that they do not overlap with each other. A criterion for such a condition is given by the so-called Le Roy radius \cite{leroy74,ji95} defined as $R_{LR}=2(\sqrt{\langle r_A^2\rangle}+\sqrt{\langle r_B^2\rangle})$, where $\langle r_A^2\rangle$ and $\langle r_B^2\rangle$ are the averaged squared distance of the outermost electron from the origin of each charge distribution $A$ and $B$, respectively. The electrostatic potential energy of interaction between $A$ and $B$ can be written as an expansion over products of multipole moments of $A$ and $B$ located at a distance $R$ from each other
\begin{eqnarray}
\hat{V}_{AB}(R) & = & \sum_{L_{A},L_{B}=0}^{+\infty}\sum_{M=-L_{<}}^{L_{<}}\frac{1}{R^{1+L_{A}+L_{B}}}\nonumber \\
 & \times & f_{L_{A}L_{B}M}\hat{Q}_{L_{A}}^{M}(\hat{r}_{A})\hat{Q}_{L_{B}}^{-M}(\hat{r}_{B})\,,
\label{eq:LR-Potential}
\end{eqnarray}
where $L_{<}=\min(L_A,L_B)$. The operator $\hat{Q}_{L_{X}}^{M}(\hat{r}_{X})$ is associated to the $2^{L_{X}}$-pole of the charge distribution $X$ $(X=A$ or $B)$, expressed in the body-fixed coordinate system with the origin at the center of mass of $X$
\begin{equation}
\hat{Q}_{L_{X}}^{M}(\hat{r}_{X})=\sqrt{\frac{4\pi}{2L_{X}+1}}\sum_{i\in X}q_{i}\hat{r}_{i}^{L_{X}}Y_{L_{X}}^{M}(\hat{\theta}_{i},\hat{\phi}_{i})\,,
\label{eq:q_tensor}
\end{equation}
where $q_i$ is the value of each charge $i$ composing the distribution $X$. The two coordinates systems (centered at $A$ and $B$) are assumed to have parallel axes with the $Z$ axis that goes from the center of mass of $A$ towards $B$ (see Fig. \ref{fig:CS}). This choice of  $Z$  implies in Eqs. (\ref{eq:LR-Potential}) and (\ref{eq:q_tensor}) that $M_A=-M_B \equiv M$, where  $M_A$ and $M_B$ are the projections of  $L_A$ and $L_B$, so that the factor $f_{L_{A}L_{B}M}$ is equal to
\begin{eqnarray}
f_{L_{A}L_{B}M} & = & \frac{\left(-1\right)^{L_{B}}\left(L_{A}+L_{B}\right)!}{\sqrt{\left(L_{A}+M\right)!\left(L_{A}-M\right)!}}\nonumber \\
 & \times & \frac{1}{\sqrt{\left(L_{B}+M\right)!\left(L_{B}-M\right)!}}\,.
 \label{eq:LR-gM-LALB}
\end{eqnarray}

The energy of interaction between the two charge distributions is calculated using perturbation theory. To the lowest (zeroth-) order of perturbation theory, the two systems are independent and the total energy is the sum of the individual energies
\begin{equation}
E_{0}^{0}=E_{A0}^{0}+E_{B0}^{0}\,,
\label{eq:LR-E0}
\end{equation}
and the total wave function is the product of individual wave functions
\begin{equation}
\left|\Psi_{0}^{0}\right\rangle =\left|\Psi_{A0}^{0}\right\rangle \left|\Psi_{B0}^{0}\right\rangle\,.
\label{eq:LR-psi0}
\end{equation}
In Eqs. (\ref{eq:LR-E0}), (\ref{eq:LR-psi0}) and below, the superscript labels the perturbation order, and the subscript labels the unperturbed states.

In the present study the system $A$ is the alkali-metal dimer and $B$ is the alkali-metal atom. We consider the dimer in a vibrational level $v_d$ of its fundamental electronic state $\left|X^{1}\Sigma_{g}^{+},v_{d}\right\rangle$, and in an arbitrary rotational state $\left|j,m_{j}\right\rangle $. In order to investigate a realistic approach for atom-molecule photoassociation, we consider the atom $B$ with a single outer electron being excited to the $p$ state $\left|n,\ell=1,\lambda\right\rangle $. However, we ignore in the following the fine structure of the excited atom for clarity, as discussed later in the text. The projections $m_j$ and $\lambda$ are defined with respect to the $Z$ axis. The energy origin corresponds to an infinite separation between the atom and the dimer. Thus, the unperturbed energy reduces to
\begin{equation}
E_{0}^{0}=B_{v_d}j\left(j+1\right)\,,
\end{equation}
where $B_{v_d}$ is the rotational constant of the dimer in its vibrational level $v_d$. The atomic state of $B$ is expressed in the $LS$ coupling case, because the operators in the interaction potential of Eq. (\ref{eq:LR-Potential}) act only on the coordinate part of wave functions. The first order correction  $E_{0}^{1}$ to the energy is due to the permanent multipoles of $A$ and $B$. In our case, both distributions exhibit a permanent quadrupole moment in their body-fixed frame, so that the most important contribution comes from the quadrupole-quadrupole interaction with an asymptotic coefficient $C_5$
\begin{equation}
E_{0}^{1}=\frac{C_{5}}{R^{5}}\,.
\end{equation}
 %Obviously, for heavy alkali-metal atoms, where spin-orbit is an important contribution to the atomic energy, the $jj$ coupling is more appropriate. The main difference with the $LS$ formulas discussed here would be the change of the atomic state from $P\equiv\left|n,\ell,\lambda\right\rangle $ to $P_j\equiv\left|n,\ell,j,\lambda_j\right\rangle$ (with $j=1/2$ or 3/2), where the state $P_j$ is written as the appropriate superposition of atomic states with different $\lambda$ and spin projections. We thus expect that the generalization of the present formalism to the $jj$ coupling scheme will multiply our $C_5$ coefficients by a factor of the order of 1.

\section{Calculation of the $C_{5}$ coefficient}
\label{sec:c5}

\begin{figure}
\begin{centering}
\includegraphics[width=0.6\textwidth]{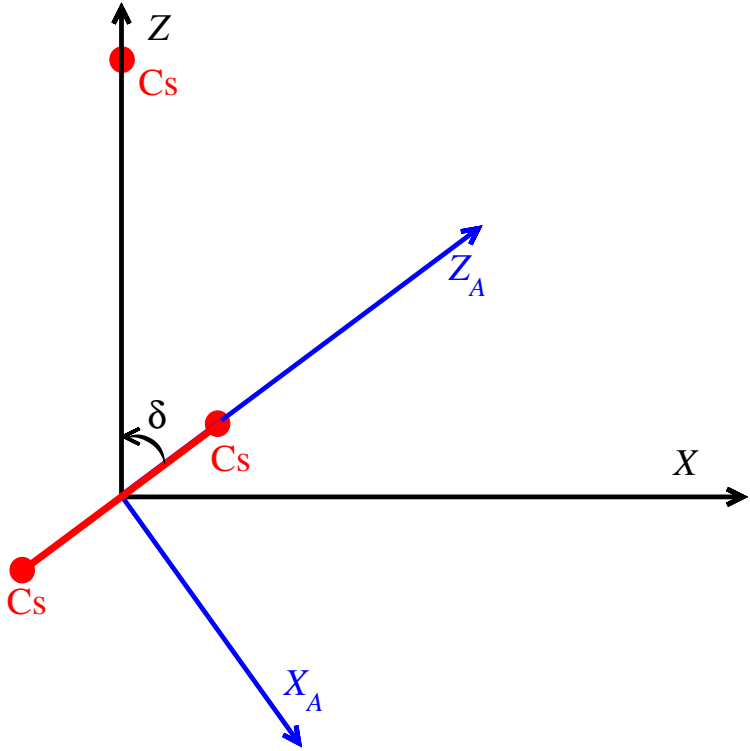}
\par\end{centering}
\caption{\label{fig:CS}The two coordinate systems, $X_AY_AZ_A$ (D-CS) and $XYZ$ (T-CS) defined for the dimer and for the trimer, respectively. The $Y$ and $Y_A$ axes coincide and point into the plane of the figure. The subsystem $A$ in this figure is the Cs$_2$ molecule, the subsystem $B$ is the Cs atom. The T-CS is related to the laboratory coordinate system $(\tilde{x} \tilde{y} \tilde{z})$ by the usual Euler angles $(\alpha,\beta,\gamma)$, not represented here.}
\end{figure}

The $C_{5}$ coefficient is calculated for arbitrary values of $m_{j}$ and $\lambda$, using the degenerate perturbation theory. We define two body-fixed coordinate systems (CS) (Fig. \ref{fig:CS}). The first CS (we call it the dimer CS, or D-CS) with axes $X_A,Y_A$, and $Z_A$ has  as the origin the center of mass of the dimer. The $Z_A$ axis is the dimer axis and the $Y_A$ axis is orthogonal to the plane of the trimer. The second CS (trimer CS, or T-CS) with axes $XYZ$ is such that the $X,Z$ axes are also (as $X_A$ and $Z_A$) in the plane of the trimer, while $Z$ is oriented from the center of the dimer towards the atom $B$; the $Y_A$ and $Y$ axes are identical. The T-CS is deduced from the D-CS by a rotation with an angle $\delta$ around the $Y$ axis.

The perturbation Hamiltonian $V_{AB}^{qq}(R)$ for the quadrupole-quadrupole interaction is given by setting $L_{A}=L_{B}=2$ in Eq. (\ref{eq:LR-Potential}):
\begin{equation}
\hat{V}_{AB}^{qq}(R)=\frac{24}{R^{5}}\sum_{M=-2}^{2}\frac{\hat{Q}_{2}^{M}(\hat{r}_{A})\hat{Q}_{2}^{-M}(\hat{r}_{B})}{\left(2+M\right)!\left(2-M\right)!}\,.
\label{eq:LR-Pot-qq}
\end{equation}
The Hamiltonian has the  form of a sum of tensor products, composed of operators acting in subspaces of the unperturbed eigenstates of $A$ and $B$. Usually, the $2^{L}$-pole tensor components $\hat{q}_{L}^{M'}$ of a charge distribution are defined in its proper CS, \textit{i.e.} $X_AY_AZ_A$ for the dimer $A$. Therefore, the $\hat{Q}_{L}^{M}$ tensor components in the T-CS are written as
\begin{equation}
\hat{Q}_{L}^{M}=\sum_{M'=-L}^{L}d_{MM'}^{L}(\delta)\hat{q}_{L}^{M'},
\label{eq:LR-rotation-QLM}
\end{equation}
where $d_{MM'}^{L}(\delta)$ are the reduced Wigner matrix elements. In the case of the alkali-metal dimer in the  $^{1}\Sigma_{g}^{+}$ state, the only non-zero component of the quadrupole moment  is $\hat{q}_{2}^{0}$ and  Eq. (\ref{eq:LR-rotation-QLM}) reduces to
\begin{equation}
\hat{Q}_{2}^{M}=d_{M0}^{2}(\delta)\hat{q}_{2}^{0}\,.
\end{equation}
The component $\hat{q}_{2}^{0}$ is just a scalar parameter, which will below be referred to as $q_{2}^{0}$.

In the T-CS, the wave function of the rotational state of the dimer $\left|jm_{j}\right\rangle$ is written as $\sqrt{\left(2j+1\right)/2}\, d_{m_{j}0}^{j}(\delta)$, depending only on the internal angle $\delta$. The normalization constant is such that the integral over angle $\delta$ is unity. We obtain the following expression for matrix elements of the operator $\hat{Q}_{2}^{M}$
\begin{eqnarray}
 \left\langle jm_{j}'\left|\hat{Q}_{2}^{M}\right|jm_{j}\right\rangle& = &  \frac{2j+1}{2}q_{2}^{0}\int_{0}^{\pi}d\delta d_{m'_{j}0}^{j}(\delta)d_{M0}^{2}(\delta)d_{m_{j}0}^{j}(\delta)\nonumber \\
 & = & C_{20j0}^{j0}C_{2Mjm}^{jm'_{j}}q_{2}^{0},
 \label{eq:LR-quadrup-molec}
\end{eqnarray}
where the Clebsch-Gordan coefficients $C_{\ell m \ell' m'}^{\ell''m''}$  appear after intergrating the product of three $d_{MM'}^{L}$ functions \cite{varshalovich88}. The zeroth-order energy $E_{0}^{0}$ depends on $j$ and is degenerate for all values of $m_{j}$. Thus, the perturbation Hamiltonian (Eq.(\ref{eq:LR-Pot-qq})) has to be evaluated with the degenerate perturbation theory. Indeed, Eq. (\ref{eq:LR-quadrup-molec}) shows that the quadrupole moment has matrix elements for different values of $m_{j}$ because of the $m'_{j}=m_{j}+M$ selection rule. The degeneracy between different $m_{j}$ values will be removed, leading to different values of $C_5$. This is the key point of the present treatment, as anisotropic values of $C_5$ are determined as functions of quantum numbers of the partners, and not restricted to a given geometry.

Assuming the alkali-metal atom being in the state labeled $\left|n\ell\lambda\right\rangle $ (the spin is neglected here), we calculate the matrix elements of the atomic quadrupole moment operator for a given $\ell$  between two different Zeeman sublevels $\lambda$ and $\lambda'$ following the same treatment as above. We obtain
\begin{eqnarray}
\left\langle n\ell\lambda'\left|\hat{Q}_{2}^{M}\right|n\ell\lambda\right\rangle  & = & -\sqrt{\frac{4\pi}{5}}\left\langle r_{n\ell}^{2}\right\rangle \int_{0}^{2\pi}d\phi \int_{0}^{\pi}d\theta Y_{\ell}^{\lambda'*}Y_{2}^{M}Y_{\ell}^{\lambda},
\label{eq:LR-quadrup-ato-1}
\end{eqnarray}
where the negative sign comes from the electron charge.  The mean squared position $\left\langle r_{n\ell}^{2}\right\rangle $  of the valence electron is independent on $\lambda$. Using the properties of spherical harmonics, we rewrite Eq. (\ref{eq:LR-quadrup-ato-1}) as
\begin{equation}
\left\langle n\ell\lambda'\left|\hat{Q}_{2}^{M}\right|n\ell\lambda\right\rangle =-C_{20\ell0}^{\ell0}C_{2M\ell\lambda}^{\ell\lambda'}\left\langle r_{n\ell}^{2}\right\rangle.
\label{eq:LR-quadrup-ato-2}
\end{equation}
The situation is analogous to the molecular case: if $M\neq0$, the operator $\hat{Q}_{2}^{M}$ couples $\lambda$ to $\lambda'=\lambda+M$, and the perturbation Hamiltonian lifts the degeneracy with respect to $\lambda$ also.

Summarizing the above results, the perturbation operator of Eq. (\ref{eq:LR-Pot-qq}) couples the $\left(2j+1\right)$ rotational states of the molecule with a given value of $j$, and the $\left(2\ell+1\right)$ Zeeman states of the atom with a given value of $\ell$. The $C_{5}$ coefficients are then given by $\left(2j+1\right)\times \left(2\ell+1\right)$ eigenvalues of the operator $\hat{V}_{AB}^{qq}$. Using Eqs. (\ref{eq:LR-quadrup-molec}) and (\ref{eq:LR-quadrup-ato-2}), the matrix elements of $\hat{V}_{AB}^{qq}$ are written
\begin{eqnarray}
 \left\langle jm'_{j}\ell\lambda'\left|V_{AB}^{qq}\right|jm_{j}\ell\lambda\right\rangle& = &  -24C_{20j0}^{j0}C_{20\ell0}^{\ell0}\frac{q_{2}^{0}\left\langle r_{n\ell}^{2}\right\rangle }{R^{5}}\nonumber \\
 & \times & \sum_{M=-2}^{2}\frac{C_{2Mjm_{j}}^{jm'_{j}}C_{2-M\ell\lambda}^{\ell\lambda'}}{\left(2+M\right)!\left(2-M\right)!}\,.
 \label{eq:LR-pot-qq-MatrElem}
\end{eqnarray}
From the integration over Euler angles and the properties of Clebsch-Gordan coefficients, the following  selection rules for $\hat{V}_{AB}^{qq}$ are derived: (1) The projection $\widetilde{m}_{J}$ of the total orbital momentum $\vec{J}=\vec{j}+\vec{\ell}$ on the laboratory $\widetilde{z}$ axis is conserved. (2) The projection $m_J=m_{j}+\lambda$ of the total orbital momentum $\vec{J}$ on the $Z$ axis of T-CS is conserved. This rule can also be deduced by the combination of Eq.~(\ref{eq:LR-quadrup-molec}) and Eq.~(\ref{eq:LR-quadrup-ato-2}).

Equation (\ref{eq:LR-pot-qq-MatrElem}) demonstrates the equivalence between the atomic orbital momentum $\vec{\ell}$ and the dimer rotation $\vec{j}$ in the formalism, which describes the long-range interaction between two charge distributions with defined angular momenta  irrespective to their internal structure. If one of the two angular momenta is zero, the corresponding quadrupole moment vanishes, and the $C_5$ coefficient as well. Therefore, the interaction will be the usual $C_6/R^{6}$ van der Waals potential. If neither of the two angular momenta, $j$ and $\ell$ is zero, the long-range interaction varies as $C_5/R^5$ and, therefore, the potential has a larger density of vibrational states close to the dissociation limit than the lowest electronic state of the system  when the atom in its ground $S$ state. Such a situation is favorable for the photoassociation of atom-molecule pairs into excited trimers, just like for the photoassociation of identical atom pairs (see for example, experimental work of Ref. \cite{pillet97}).

\section{Results and discussion}
\label{sec:results}

To illustrate the previous formalism, we first consider analytically the simplest case $j=\ell=1$. The perturbation Hamiltonian $V_{AB}^{qq}$ reduces to a $9\times9$ matrix with elements calculated from tensor products of the atomic and dimer states. For simplicity, we omit the $j$ and $\ell$ labels in the following, and the quantum states of the atom-molecule pair are denoted by projections $\left\{ \left|m_{j},\lambda\right\rangle \right\}$ only. All such states form the basis of the representation. If we sort the states by values of the conserved projection of the angular momentum $m_{J}=m_{j}+\lambda$, we obtain the matrix of $V_{AB}^{qq}$ in a block-diagonal form.

The two blocks defined by $\left|m_{j},\lambda\right\rangle =\left|-1,-1\right\rangle$ and $\left|+1,+1\right\rangle$ ($m_J=\pm2$) reduce to a single element with a negative value of the corresponding coefficient
\begin{equation}
C_5=-\frac{6q_{2}^{0}\left\langle r_{n\ell=1}^{2}\right\rangle }{25}\,.
\end{equation}
It produces an attractive interaction. Two other $2\times 2$ blocks (with $m_J=\pm1$) are defined by the two subspaces $\left\{ \left|-1,0\right\rangle ;\left|0,-1\right\rangle \right\} $ and $\left\{ \left|0,1\right\rangle ;\left|1,0\right\rangle \right\} $. The corresponding $C_{5}$ coefficients are $\frac{24q_{2}^{0}\left\langle r_{n\ell}^{2}\right\rangle }{25}$ (positive value) and zero. Finally, the last $3\times3$  block comes from the subspace $\left\{ \left|-1,1\right\rangle ;\left|0,0\right\rangle ;\left|1,-1\right\rangle \right\} $ ($m_J=0$). Two of the corresponding $C_{5}$ coefficients are zero, and the third one is
\begin{equation}
C_{5}=-\frac{36q_{2}^{0}\left\langle r_{n\ell=1}^{2}\right\rangle }{25},
\label{eq:LR-C5min}
\end{equation}
with the eigenvector $\frac{1}{\sqrt{6}}\left(\left|-1,1\right\rangle +2\left|0,0\right\rangle +\left|1,-1\right\rangle \right)$.
The coefficient in Eq. (\ref{eq:LR-C5min}) is negative with the largest magnitude  out of all $C_5$ coefficients obtained in the case of $j=\ell=1$. It corresponds to the most attractive configuration between the atom and the dimer, and is expected to be the most favorable for the photoassociation.

\begin{table}
\begin{centering}
\begin{tabular}{|c|c|c|c|}
\hline
\noalign{\vskip\doublerulesep}
$m_J$&$\left|\Phi_{0}^{0}\right\rangle $ & $C_{5}$ $\left(q_{2}^{0}\left\langle r_{n\ell}^{2}\right\rangle \right)$ & $C_{5}$ (a.u.)\tabularnewline[\doublerulesep]
\hline
\noalign{\vskip\doublerulesep}
-2&$\left|-1,-1\right\rangle $ & $-\frac{6}{25}$ & -279 \tabularnewline[\doublerulesep]
\noalign{\vskip\doublerulesep}
-1&$\frac{1}{\sqrt{2}}\left(\left|-1,0\right\rangle +\left|0,-1\right\rangle \right)$ & $\frac{24}{25}$ & 1116 \tabularnewline[\doublerulesep]
\noalign{\vskip\doublerulesep}
-1&$\frac{1}{\sqrt{2}}\left(\left|-1,0\right\rangle -\left|0,-1\right\rangle \right)$ & 0 & 0\tabularnewline[\doublerulesep]
\noalign{\vskip\doublerulesep}
0&$\frac{1}{\sqrt{6}}\left(\left|-1,1\right\rangle +2\left|0,0\right\rangle +\left|1,-1\right\rangle \right)$ & $-\frac{36}{25}$ & -1674 \tabularnewline[\doublerulesep]
\noalign{\vskip\doublerulesep}
0&$\frac{1}{\sqrt{3}}\left(\left|-1,1\right\rangle -\left|0,0\right\rangle +\left|1,-1\right\rangle \right)$ & 0 & 0\tabularnewline[\doublerulesep]
\noalign{\vskip\doublerulesep}
0&$\frac{1}{\sqrt{2}}\left(\left|-1,1\right\rangle -\left|1,-1\right\rangle \right)$ & 0 & 0\tabularnewline[\doublerulesep]
\noalign{\vskip\doublerulesep}
+1&$\frac{1}{\sqrt{2}}\left(\left|1,0\right\rangle -\left|0,1\right\rangle \right)$ & $\frac{24}{25}$ & 1116 \tabularnewline[\doublerulesep]
\noalign{\vskip\doublerulesep}
+1&$\frac{1}{\sqrt{2}}\left(\left|1,0\right\rangle +\left|0,1\right\rangle \right)$ & 0 & 0\tabularnewline[\doublerulesep]
\noalign{\vskip\doublerulesep}
+2&$\left|1,1\right\rangle $ & $-\frac{6}{25}$ & -279 \tabularnewline[\doublerulesep]
\hline
\end{tabular}
\par\end{centering}
\begin{centering}
\caption{\label{tab:LR-C5-j1}Values of the $C_{5}$ coefficient and their corresponding eigenvectors characterized by their $m_J$ value, for Cs$_2(X^1\Sigma_g^+, v_d=0, j=1)$+Cs($6P$). The values of $C_{5}$ are given in units of $q_{2}^{0}\left\langle r_{n\ell}^{2}\right\rangle $ in the second column, and in atomic units for Cs$_{2}+$Cs in the third column. For cesium, the data are: $\left\langle r_{6P}^{2}\right\rangle =62.65$ a.u. and $q_{2}^{0}=18.56$ a.u. (see text). Due to the uncertainty over $q_2^0$, the results are given with a precision of 1 a.u.}
\par\end{centering}
\end{table}

The results of the calculation for the case  $j=\ell=1$ are summarized in Table \ref{tab:LR-C5-j1}. The second column of the table gives the eigenvectors $\left|\Phi_{0}^{0}\right\rangle $ of the Hamiltonian of Eq. (\ref{eq:LR-Pot-qq}) in the $j=1$ subspace. The third column gives the so-called \textit{reduced} values of  $C_{5}$ in units of $q_{2}^{0}\left\langle r_{n\ell}^{2}\right\rangle $, which stress the general character of our treatment: it can be applied to all alkali-metal trimers, but it can also be compared with the existing results on the long-range interaction between two excited atoms \cite{marinescu97}. The eigenvectors obtained here are the same as in Ref. \cite{marinescu97}, but the signs of the $C_{5}$ coefficients are opposite to the coefficients obtained in Ref. \cite{marinescu97}. The reason is clear from Eqs. (\ref{eq:LR-quadrup-molec}) and (\ref{eq:LR-quadrup-ato-1}): The signs of the matrix elements of quadrupole moments for the dimer Eq. (\ref{eq:LR-quadrup-molec}) and the atom Eq. (\ref{eq:LR-quadrup-ato-1}) are opposite. When they are combined together in Eq. (\ref{eq:LR-Pot-qq}) they give an additional negative sign to the perturbation. The two matrix elements of the atomic quadrupoles give the positive sign to the perturbation matrix elements.

The fourth column in Table \ref{tab:LR-C5-j1} displays estimates for the $C_{5}$ coefficients for Cs$_{2}+$Cs. To the best of our knowledge, there are no available values for the quadrupole moment of Cs$_{2}$ in the literature. Therefore, we calculated it for the electronic ground state with the Gaussian quantum chemistry package \cite{gaussian03} using the MP2 method with the Def2-TZVPP basis \cite{weigend05}. To check the accuracy of such an estimation, we first calculated the quadrupole moment of $\textrm{K}_2$ and compared it to available accurate \textit{ab initio} calculations \cite{harrison05}. We obtained 12.258 a.u., which differs by a factor 1.28 from the value 15.689 a.u. of Ref. \cite{harrison05}.  For cesium, the Def2-TZVPP basis  \cite{weigend05} contains also effective core potentials (ECPs) standing for the 54 inner electrons of the core. We obtained for  $\textrm{Cs}_2$ the value of 14.51 a.u. that we multiplied by the same factor to estimate the Cs$_2$ quadrupole moment to $q_{2}^{0}=18.58\textrm{ a.u.}$ The mean squared radius of the $6P$ orbital of cesium, which is 62.65 a.u., is calculated using a Dirac-Fock method  \footnote{M. Aymar, private communication}. It is worth to mention that the  values of $C_{5}$  shown in the table are of the same order of magnitude as the values for Cs($6P$)+Cs($6P$)  \cite{marinescu97}.

For $\ell=1$ and arbitrary $j$, the perturbation Hamiltonian is a $3(2j+1)\times3(2j+1)$ matrix, that can be diagonalized numerically. The eigenvalues obtained numerically for $j=2$ to 4 are given in Table \ref{tab:LR-C5-j-geq-2}. The $C_{5}$ coefficients are of the same order of magnitude as for $j=1$, but in average they become smaller in magnitude as $j$ increases, due to smaller Clebsch-Gordan coefficients. The $C_5$ coefficients are sorted by values of $\left|m_J\right|$, which, in analogy to diatomic molecules, are labeled $\Sigma$, $\Pi$, $\Delta$, $\Phi$, $\Gamma$, and H for $\left|m_J\right|=0$ to 5. For $\Sigma$ states, the reflection symmetry through the $Z$ axis is also considered, giving the usual +/- superscripts. For states other than $\Sigma$ the sign +/- is not specified because such states are degenerate (in the present approximation) with respect to the reflection.

\begin{table}
\begin{centering}
\begin{tabular}{|c|c|c||c|c|c|}
\hline
symmetry & ~~$j$~~ & $C_{5}$ (a.u.) & symmetry & ~~$j$~~ & $C_{5}$ (a.u.)\tabularnewline
\hline
\hline
%1a
$\Sigma^{+}$ & 2 & -913  &
%19a
$\Delta$     & 2 & -140  \tabularnewline
%2a
          & 2 &  116  &
%20a
          & 2 & 1136  \tabularnewline
%3a
          & 3 & -796  &
%1b
          & 3 & -835  \tabularnewline
%4a
          & 3 &  145  &
%2b
          & 3 & -87   \tabularnewline
%5a
          & 4 & -755  &
%3b
          & 3 &  736  \tabularnewline
%6a
          & 4 &  157  &
%4b
          & 4 & -721  \tabularnewline
%7a
$\Sigma^{-}$ & 2 &  399  &
%5b
          & 4 & -11   \tabularnewline
%8a
          & 3 &  465  &
%6b
          & 4 &  623  \tabularnewline
%9a
          & 4 &  489  &
%7b
$\Phi$   & 2 & -399  \tabularnewline
%10a
$\Pi$        & 2 & -964  &
%10b
          & 3 & -245  \tabularnewline
%11a
          & 2 & -19  &
%11b
          & 3 & 1175  \tabularnewline
%12a
          & 2 &  584  &
%13b
          & 4 & -783  \tabularnewline
%13a
          & 3 & -783  &
%14b
          & 4 & -161  \tabularnewline
%14a
          & 3 &  64  &
%15b
          & 4 &  835  \tabularnewline
%15a
          & 3 &  532  &
%16b
$\Gamma$ & 3 & -465  \tabularnewline
%16a
          & 4 & -739  &
%19b
          & 4 & -320  \tabularnewline
%17a
          & 4 &  108  &
%20b
          & 4 & 1208  \tabularnewline
%18a
          & 4 &  522  &
%21b
H & 4 & -507 \tabularnewline\hline
\end{tabular}
\par\end{centering}
\caption{\label{tab:LR-C5-j-geq-2} The $C_{5}$ coefficients of the Cs$_2(X^1\Sigma_g^+, v_d=0, j)$+Cs($6P$) long-range interaction calculated numerically for $j=2$ to 4.  $C_5$ are sorted by projections $m_J=m_j+\lambda$ of the total orbital angular momentum  on the $Z$ axis, and by the sign +/- of the wave function with respect to a reflection through the plane containing the $Z$ axis. In analogy to a diatomic molecule, the eigenstates are labeled with $\Sigma^{+/-}, \Pi, \Delta, \Phi, \Gamma$, and H for $m_J=0, 1, 2, 3, 4, 5$, respectively. The values for $\left\langle r_{6P}^{2}\right\rangle $ and $q_{2}^{0}$ are the same as in Table \ref{tab:LR-C5-j1}.}
\end{table}

\begin{figure}
\begin{centering}
\includegraphics[width=15cm]{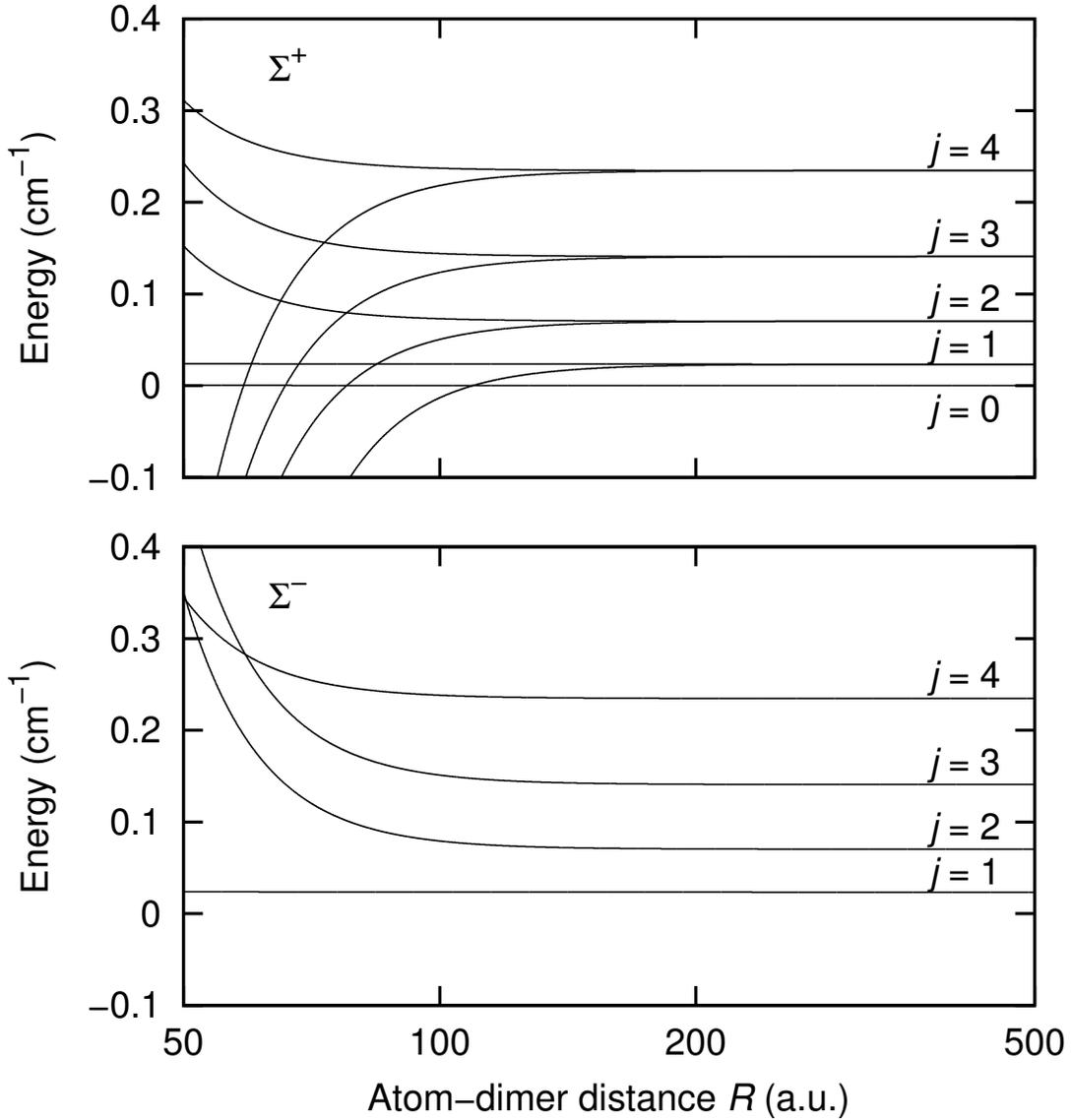}
\par\end{centering}
\caption{Long-range potential energy curves $C_{5}/R^{5}$ as a function of the atom-dimer distance $R$ (notice logarithmic scale along $R$), for the $\Sigma^+$ and $\Sigma^-$ symmetries, and for the five lowest rotational levels of Cs$_{2}(X^1\Sigma_g^+(v_d=0))$. The curves are drawn for distances larger than the Le~Roy radius $R_{LR}=45$~a.u.}
\label{fig:C5_Ener_R-Sigma}
\end{figure}

\begin{figure}
\begin{centering}
\includegraphics[width=15cm]{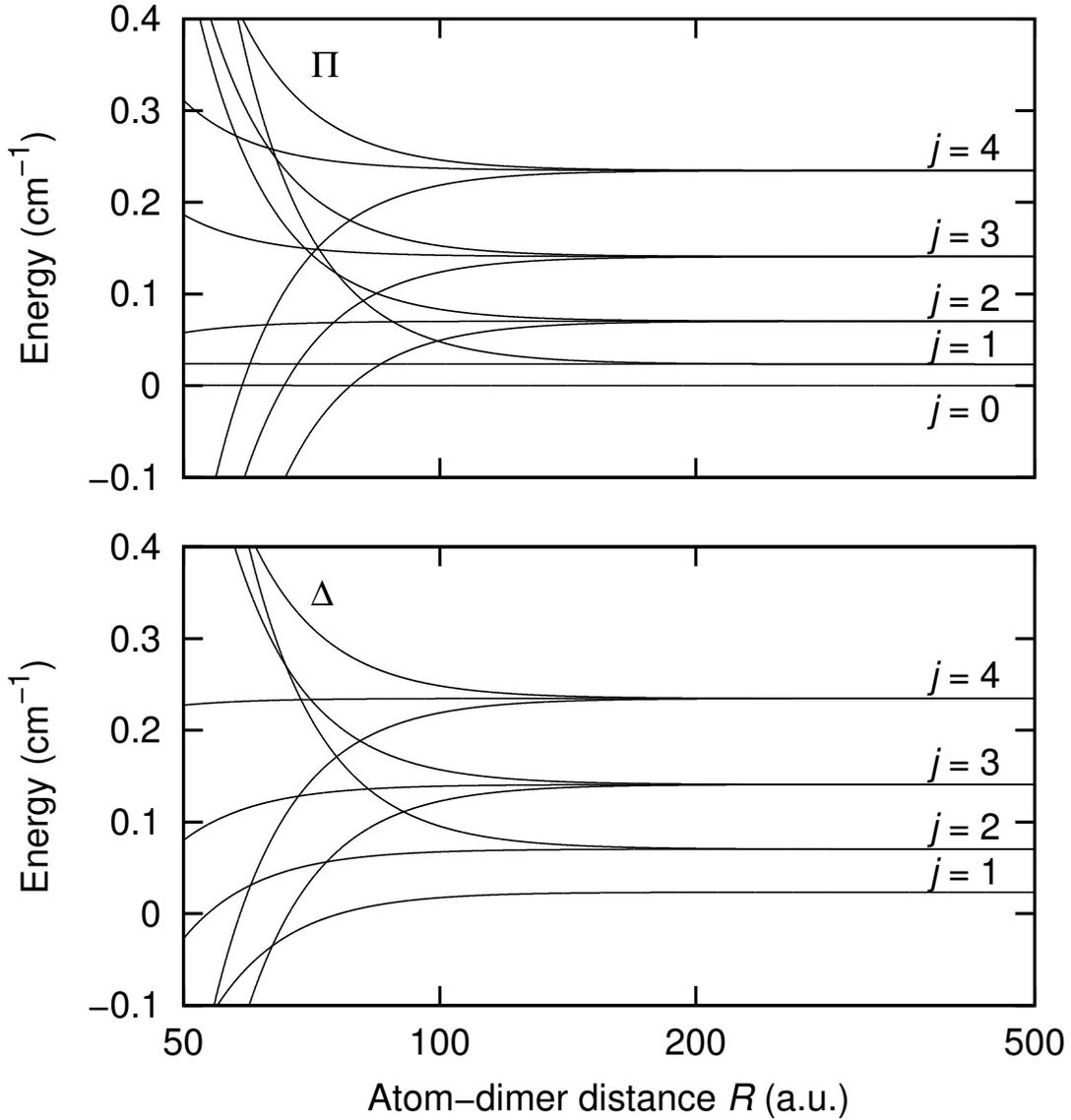}
\par\end{centering}
\caption{Same as Fig. \ref{fig:C5_Ener_R-Sigma} for $\Pi$ and $\Delta$ symmetries.}
\label{fig:C5_Ener_R-Pi-Delta}
\end{figure}

\begin{figure}
\begin{centering}
\includegraphics[width=15cm]{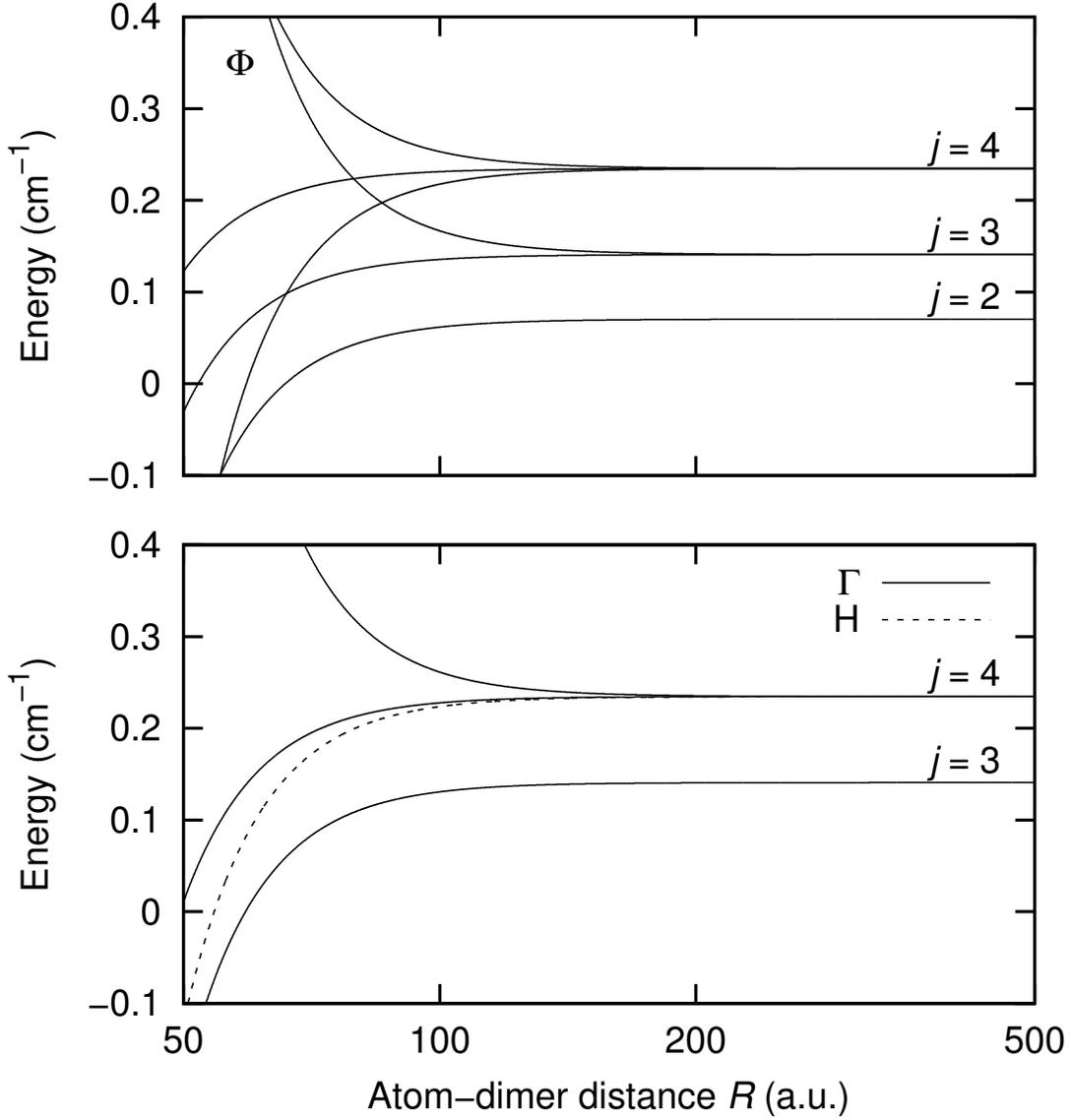}
\par\end{centering}
\caption{Same as Fig. \ref{fig:C5_Ener_R-Sigma} for $\Phi$, $\Gamma$ and H symmetries.
The upper panel displays $\Phi$ symmetry, and the lower one displays both $\Gamma$ symmetry (solid lines) and H symmetry (dashed lines).}
\label{fig:C5_Ener_R-Phi-Gamma}
\end{figure}

We use the same symmetry notations in Figs. \ref{fig:C5_Ener_R-Sigma}, \ref{fig:C5_Ener_R-Pi-Delta} and \ref{fig:C5_Ener_R-Phi-Gamma}, where we display the long-range potential energy curves $C_{5}/R^{5}$ for the Cs$_2$+Cs($6P$) system calculated for the first five rotational levels $j$ of Cs$_2$ as a function of the atom-dimer distance $R$. The energies of dissociation are given by Cs$_2$ rotational energies, $B_{0}j\left(j+1\right)$, $j=0,\dots,4$. The rotational constant for ground vibrational level of Cs$_{2}$ is $B_{0}=1.17314\times10^{-2}$ cm$^{-1}$ \cite{amiot02}.

The potential energy curves are shown up to $R=500$~a.u. Beyond this limit, the distance between Cs$_{2}$ and Cs becomes comparable to the wavelengths of relevant atomic and molecular transitions, which are in the optical frequency domain. Therefore, in that region, electrodynamics effects, for example, due to retardation, should be taken into account \cite{casimir48}.

As already mentioned, the lower limit of the region where the present approximation is applicable can be estimated by the Le Roy radius $R_{LR}=2\left(\sqrt{\langle r_{0}^{2}(\textrm{Cs}_{2})\rangle}+\sqrt{\langle r_{6P}^{2}(\textrm{Cs})\rangle}\right)$, where $\langle r_{0}^2\rangle$ and $\langle r_{6P}^2\rangle$ are related the extension of the dimer and atomic electronic clouds, respectively. For the atom, one has $\left\langle r_{6P}^{2}\right\rangle=62.65$ a.u., which is given in Table \ref{tab:LR-C5-j1}. As for the dimer, $\langle r_{0}^2\rangle$ is calculated from the elements of the quadrupole tensor. All its non-diagonal elements $Q_{\alpha\beta}$ are zero for $^1\Sigma_g^+$ states in D-CS. Its diagonal elements $Q_{\alpha\alpha}$ ($\alpha=X_A,Y_A$, or $Z_A$) in D-CS are estimated using the Gaussian package, and is formally written as a sum over all charges
\begin{equation}
Q_{\alpha\alpha} = \sum_{i} q_i \left\langle \alpha_i^2 \right\rangle.
\end{equation}
%
%For an approximate estimation of the dimer quadrupole moment, we can consider that closed electronic shells of each of the two Cs atoms are not perturbed by the presence of the other atom. It means that the quadrupole moment of the dimer is made only by valence electrons of the dimer.
As the two nuclei of Cs$_2$ are along the $Z_A$ axis and it is a $^1\Sigma_g^+$ molecular state, $Q_{X_A X_A}$ is equal to $Q_{Y_A Y_A}$ and both are functions of the coordinates of two valence electrons only ($i=1, 2$)
\begin{equation}
Q_{X_A X_A} = -e\sum_{i=1}^{2} \left\langle X_{Ai}^2 \right\rangle\,.
\end{equation}
Now considering for simplicity that the cores are fixed at the position $Z_A=\pm{r_e}/2$ (the rigid rotor approximation, valid for $v_d=0$), for $Q_{Z_A Z_A}$ we obtain
\begin{equation}
Q_{Z_A Z_A} = e\frac{r_e^2}{2}-e\sum_{i=1}^{2} \left\langle Z_{Ai}^2 \right\rangle.
\end{equation}
Now, setting
\begin{equation}
\langle r_0^2 \rangle = \sum_{i=1}^2 \sum_{\alpha =(X_A,Y_A,Z_A)} \left\langle \alpha_i^2 \right\rangle,
\end{equation}
we reach the final expression for $\langle r_0^2 \rangle$
\begin{equation}
\langle r_0^2 \rangle = \frac{r_e^2}{2}-Q_{Z_A Z_A}-2Q_{X_A X_A}\,,
\label{R-LR-quadrup}
\end{equation}
where $e=1$ in atomic units. The $Q_{\alpha\alpha}$ matrix elements are calculated with the same method as $q_2^0$ (using the ratio 1.27 to the K$_2$ value) \footnote{The $q_2^0$ and $Q_{\alpha\alpha}$ elements are connected to each other by $q_2^0=2Q_{Z_A Z_A}-Q_{X_A X_A}-Q_{Y_A Y_A}$.}, which yields $Q_{X_A X_A}=-69$~a.u., $Q_{Z_A Z_A}=-41$~a.u. and $r_e=8.7$~a.u. Therefore, we obtain $\langle r_0^2 \rangle=216$~a.u. and the Le~Roy radius $R_{LR}=45$~a.u.

As we can see from Figs. \ref{fig:C5_Ener_R-Sigma}, \ref{fig:C5_Ener_R-Pi-Delta} and \ref{fig:C5_Ener_R-Phi-Gamma}, the Le~Roy radius is smaller than the distance at which the curves start to cross each other. This is the second main result of the paper. Unlike the case of two atoms, the rotational structure of the dimer is small enough to compete with the quadrupole-quadrupole interaction. The lower limit of $R_m$ for the applicability of the present perturbation approach is thus fixed by the crossing of the potential energy curves. In order to estimate $R_m$, we note that the first crossing occurs between the curves dissociating towards to the $j=0$ limit and from the most attractive curve corresponding to $j=1$. Putting $2B_0 \equiv C_{5}^{m}/R_{m}^5$ yields a general estimate for $R_{m}$ (see Eq. (\ref{eq:LR-C5min}))
\begin{eqnarray}
R_{m} & \sim & \left(\frac{C_{5}^{m}}{2B_{0}}\right)^{1/5}=\left(\frac{18q_{2}^{0}\left\langle r_{n\ell}^{2}\right\rangle }{25B_{0}}\right)^{1/5}\nonumber \\
 & \approx & 0.936\times\left(\frac{q_{2}^{0}\left\langle r_{n\ell}^{2}\right\rangle }{B_{0}}\right)^{1/5}.\label{eq:LR-Rm}
\end{eqnarray}
For cesium, Eq.(\ref{eq:LR-Rm}) yields $R\approx102$~a.u. In the rigid rotor approximation with $B_{0}=1/\left(2\mu r_{e}^{2}\right)$ and where $\mu$ is the reduced mass of the dimer, Eq.(\ref{eq:LR-Rm})  shows that the value of $R_{m}$ is smaller for lighter atoms. For example, replacing in our treatment Cs by $^{6}$Li with the atomic parameters $\left\langle r_{2P}^{2}\right\rangle =32.5$~a.u. \cite{pipin92}, $r_{e}=5.05$~a.u. and $q_{2}^{0}=10.7$~a.u \cite{harrison05}, we obtain $R_{m}=43$~a.u. This value of $R_m$ is larger than the Le~Roy radius for lithium, for which we obtained 26~a.u. using  Eq. (\ref{R-LR-quadrup}).

For distances such that $R_{LR}<R<R_{m}$, the long-range potential (\ref{eq:LR-Potential}) is still valid, but not the perturbative approach. The non-adiabatic interaction at the curve crossings (for a given symmetry) is expected to be strong. In particular, the interaction between permanent quadrupoles would couple the dimer rotational level $j$ with $j'=j\pm2$, $j\pm4$ near the crossings. Higher-order contributions in $1/R$ should also be considered.

The number $N$ of partial waves  involved in the atom-dimer collisions depends on the temperature in the actual experiment. In order to give an upper bound for $N$, we consider a potential curve with the most attractive $C_{5}$, given by Eq. (\ref{eq:LR-C5min}) and with the added centrifugal term. It is straightforward to show that the height of the potential barrier $E_{N}^{max}$ for a given $N$ is
\begin{eqnarray}
E_{N}^{max} & = & \frac{9}{20}\left(\frac{5}{24}\right)^{2/3}\left(\frac{N\left(N+1\right)}{m}\right)^{5/3}\left(q_{2}^{0}\left\langle r_{n\ell}^{2}\right\rangle \right)^{-2/3}\nonumber \\
 & \approx & 0.158\times\left(\frac{N\left(N+1\right)}{m}\right)^{5/3}\left(q_{2}^{0}\left\langle r_{n\ell}^{2}\right\rangle \right)^{-2/3}.
\end{eqnarray}
Converted to the temperature, $E_{1}^{max}$ is approximately 1 $\mu$K for cesium. If we take typical temperatures 10-100 $\mu$K for which photoassociation experiments are achieved, only a few partial waves (6$\sim$7 for the present case) will play a significant role in the collision. This contrasts with the PA of identical atom pairs, interacting with a long-range $R^{-3}$ potential which allow much more partial waves than here.

\section{Conclusions and perspectives}

In this article, we used the multipolar expansion to calculate the long-range interaction energy of a diatomic molecule in its electronic ground state and in an arbitrary rovibrational level, and an excited atom. We applied our treatment to the case of a ground state Cs$_2$ molecule and an excited Cs$(6P)$ atom, as a prospect for cold atom/molecule photoassociation. The dimer and the atom interact through their permanent quadrupole moment. In contrast with previous works, the anisotropic interaction is computed for arbitrary geometries of the atom/molecule pair, and depends on their internal quantum numbers. We showed that the interaction lifts the degeneracy over their respective magnetic sublevels. Using the degenerate perturbation theory, we calculated the $C_5$ coefficients characterizing the quadrupole-quadrupole interaction, for the five lowest rotational levels of the ground state dimer. The photoassociation of a ground state $X^1\Sigma_g^+$ alkali-metal dimer molecule with a ground state $nS_{1/2}$ alkali-metal atom is found possible by exciting the dimer-atom system with a laser frequency red-detuned  from the $nS\to nP$ atomic transitions.

We demonstrated that the small-$R$ limit of applicability of our treatment is not due to the overlap of the electronic clouds of the partners as in the atom-atom case, but to the competition between the rotational energy of the dimer with the long-range quadrupole-quadrupole interaction. This induces crossings between potential energy curves corresponding to different rotational levels. In the region $50\le R\le 100$~a.u., the multipolar expansion is still valid, but not the perturbation approach. The inclusion of non-adiabatic couplings is required in this region for an appropriate description of the long-range behavior, as well as higher-order effects in $1/R$. This will be discussed ind detail a forthcoming paper.

It is important to stress that the above treatment has been developed in the framework of the $LS$ coupling case, in order to keep our description simple. The next step is to account for the fine structure of the excited atom. The main difference with the formulas reported here will be the change of the atomic state from $P\equiv\left|n,\ell,\lambda\right\rangle $ to $P_j\equiv\left|n,\ell,j,\lambda_j\right\rangle$ (with $j=1/2$ or 3/2), where the state $P_j$ is written as the appropriate superposition of atomic states with different $\lambda$ and spin projections through a unitary transformation. We note however that for most of the alkali-metal atoms (from Na to Cs) the fine structure is much larger than the magnitude of the long-range atom-dimer interaction. Therefore, the related $C_5$ coefficients will result from linear combinations of the coefficients of Table \ref{tab:LR-C5-j-geq-2}, and will not modify the main statement of our study concerning the range of validity of our approach. In contrast, the case of Lithium atom will be remarkable as its small fine structure of 0.335~cm$^{-1}$ \cite{radziemski95} falls within the range of energies displayed in Figs \ref{fig:C5_Ener_R-Sigma} to \ref{fig:C5_Ener_R-Phi-Gamma}, and is expected to modify the present conclusions. This work is currently under progress. A similar discussion obviously holds for the hyperfine interaction of the excited atom, which will induce even more complexity in the formalism. It could safely be neglected for all species except for Cesium (the hyperfine splitting of the $6P_{1/2}$ level is 1.167688(81) GHz \cite{udem99}.

The present formalism can be generalized to photoassociation of dipolar dimers and atoms, like KRb with K or KRb with Rb. If the atom is in the excited state $nP$ and the heteronuclear dimer in an excited rotational state $j>0$, the long-range dimer-atom interaction is dominated by a dipole-quadrupole term varying as $C_4/R^4$. The long-range interaction between two identical dipolar molecules can also be treated in the same way, as the leading term will be the usual van der Waals $C_6/R^6$ term if both molecules are in their lowest rotational level $j=0$, or dipole-dipole $C_3/R^3$ term if one of them is rotationally excited. One could thus investigate the photoassociation of two identical heteronuclear KRb ground state molecules in their lowest vibrational level, by a laser field with a frequency red-detuned with respect to $j=0\to j=1$ transition. In this respect, photoassociation of two dimers is very similar to the photoassociation of two identical alkali-metal atoms, except that the laser frequency is much smaller for the two-dimer photoassociation.

\section*{Acknowledgments}

V. K. and O. D. gratefully acknowledge stimulating discussions with Fran\c coise Masnou at the initial stage of this work. The authors are also grateful to Mireille Aymar for providing us with the values of $\left\langle r_{6P}^2 \right\rangle$ for Cs. M. L. thanks Manuel Goubet (Lab. Phlam, Univ. Lille 1) for his precious help with Gaussian. This work was done with the support of Triangle de la Physique under contract 2008-007T-QCCM (Quantum Control of Cold Molecules), and of National Science Foundation under grant PHY-0855622.

%\bibliography{DR}
%Merlin.mbs v4.21 2009-07-09.
%

\end{document}